\documentclass[preprints,article,accept,moreauthors,pdftex]{Definitions/mdpi} 

\firstpage{1} 
\pubvolume{1}
\issuenum{1}
\articlenumber{0}
\pubyear{2026}
\copyrightyear{2026}
\datereceived{} 
\dateaccepted{} 
\datepublished{} 
\hreflink{https://doi.org/} 

\usepackage{xcolor}

\usepackage{todonotes}
\usepackage{caption,subfigure}
\usepackage{bm}

\Title{Equation of State of Highly Asymmetric Neutron-Star Matter from Liquid Drop Model and Meson Polytropes}
\TitleCitation{Equation of State of Highly Asymmetric Neutron-Star Matter from Liquid Drop Model and Meson Polytropes}

\Author{Elissaios Andronopoulos $^{1}$, Konstantinos N. Gourgouliatos $^{1}$\orcidA}

\AuthorNames{Elissaios Andronopoulos}
\AuthorNames{Konstantinos N. Gourgouliatos}

\AuthorCitation{Andronopoulos, E., Gourgouliatos, K.N.}

\address{%
$^{1}$ \quad Laboratory of Universe Sciences, Department of Physics, University of Patras, 26504 Patras, Greece}

\corres{Correspondence: E. Andronopoulos, up1102087@ac.upatras.gr, K.N. Gourgouliatos kngourg@upatras.gr}

\abstract{We present a unified description of dense matter and neutron-star
structure based on simple but physically motivated models. Starting from the
thermodynamics of degenerate Fermi gases, we construct an equation of state for
cold, catalyzed matter by combining relativistic fermion statistics with the
liquid drop model of nuclear binding. The internal stratification of matter in the outer crust is described by $\beta$-equilibrium, neutron drip and a gradual transition to supranuclear matter. Short-range
repulsive interactions inspired by Quantum Hadrodynamics are incorporated at high densities in
order to ensure stability and causality.
The resulting equation of state is used as input to the
Tolman--Oppenheimer--Volkoff equations, yielding self-consistent neutron-star
models. We compute macroscopic stellar properties including the mass-radius
relation, compactness and surface redshift that can be compared with recent observational data. Despite the simplicity of the underlying microphysics, the
model produces neutron-star masses and radii compatible with current
observational constraints from X-ray timing and gravitational-wave measurements.
This work demonstrates that physically transparent models can already capture
the essential features of neutron-star structure and provide valuable insight
into the connection between dense-matter physics and astrophysical observables while they can also be used as easy to handle
models to test the impact of more complicated phenomena and variations in neutron stars. }


\begin{document}


\section{Introduction}

Compact stellar objects provide a unique natural laboratory for studying matter
under extreme conditions \cite{2016ARA&A..54..401O,2018RPPh...81e6902B,2025Univ...11..276L}. In neutron stars, densities span over ten orders of magnitude, from the outer crustal layers to the
supranuclear core, where the density exceeds that of an atomic nucleus. In this
regime, quantum degeneracy, special relativity, nuclear interactions, and general
relativity all become essential ingredients of a consistent physical
description \cite{Shapiro:1983,Haensel:2007,Glendenning:2000,potekhin2010physics}. While these questions arise from fundamental physics, recent breakthroughs in neutron star measurements of masses and radii have allowed us to probe these theoretical predictions \cite{2016RvMP...88b1001W}.

A detailed microscopic description of dense nuclear matter remains an open and
challenging problem, with many-body calculations depending sensitively on complex nuclear interactions at high density, which are intensively researched \cite{2018RPPh...81e6902B}. Nevertheless,  considerable
physical insight can be gained from simpler, phenomenological models \cite{2004AmJPh..72..892S,2006EJPh...27..577S}.

The inner structure of compact stars is determined by the balance between
pressure and gravity, which in the relativistic regime is governed by the
Tolman--Oppenheimer--Volkoff equations \cite{1939PhRv...55..364T,1939PhRv...55..374O}. Once an equation of state is specified,
these equations uniquely determine the radial profiles of pressure, density
and enclosed mass, as well as macroscopic observables such as the stellar mass and radius. Additional quantities like surface redshift and apparent radius provide direct links
between microphysical models of dense matter and astrophysical observations.

The purpose of this work is to present a simple yet physically consistent
framework for modeling neutron-star structure across a wide range of densities.
We begin by establishing the statistical mechanics of degenerate fermions,
accounting for the Fermi pressure of electrons and neutrons in various
density regimes. Next, we incorporate the liquid drop model (LDM) to describe
nuclei in the crust, capturing the dominant volume, surface, Coulomb, and
symmetry energy contributions. Using these building blocks, we construct
phenomenological polytropic approximations of the pressure at high densities,
which allow for a smooth transition from the crust to the core. 

We then extend this approach with a relativistic mean-field (RMF) treatment
of nuclear interactions, including scalar and vector meson contributions
($\sigma$, $\omega$, $\rho$), and account for the reduction of the effective
neutron mass due to the scalar field. By calibrating the meson couplings and
density-dependent scalar fractions, we obtain an equation of state that remains causal and stable up to supranuclear densities.
Finally, the resulting EOS is confronted with observational constraints,
including neutron-star masses, radii, gravitational redshifts and waves, and apparent
radii, providing a comprehensive yet computationally tractable model for
neutron-star structure \cite{Lattimer:2007,2018PhRvL.121p1101A,2025Univ...11..276L}.
  
The equation of state constructed here is explicitly phenomenological and is intended as a minimal, physically transparent baseline rather than a microscopic description of dense QCD matter. Our goal is not to model exotic degrees of freedom such as hyperons, deconfined quarks, or condensates, but to provide a unified framework in which the consequences of well-established physics—degeneracy pressure, $\beta$-equilibrium, saturation, and short-range repulsion—can be systematically explored. In this sense, the model serves as a testbed for assessing how macroscopic neutron-star observables depend on the stiffness of the EOS, largely independent of detailed microphysics.

From the perspective of high-energy and nuclear physics, this phenomenological character reflects an effective, coarse-grained description in which microscopic QCD degrees of freedom are integrated out, leaving a small set of parameters constrained by symmetry, thermodynamic consistency, and observations. The model is therefore not intended as a first-principles treatment of dense QCD matter, but as a framework appropriate for neutron-star interiors.

The structure of the paper is as follows. In Section \ref{sec:Theory}, we derive the theoretical equations that describe the system from first principles. We present the results of the equations for neutron stars in Section \ref{sec:neutron_stars}. We conclude in Section \ref{sec:Conclusions}. 

\section{Theoretical Background}
\label{sec:Theory}

Here we provide a comprehensive discussion of the theoretical considerations that will be used to derive the equations for the system. Thus, we present the basic theoretical ingredients, that will be later used in the context of neutron stars.  

\subsection{Statistical Mechanics of Degenerate Matter}

The thermodynamic properties of matter in compact stars are governed by fermion thermodynamics and statistical physics. The first law of thermodynamics reads
\begin{equation}
dU = T\,dS - P\,dV + \mu\,dN\,,
\end{equation}
here, $U$ denotes the internal energy, $S$ the entropy, $T$ the temperature, $P$ the pressure, $V$ the volume, $\mu$ the chemical potential, and $N$ the particle number of the system.  At the densities of interest, the Fermi energy of the constituents is
much larger than the thermal energy, and matter can be accurately described in
the zero-temperature limit \cite{1961ApJ...134..669S,1961ApJ...134..683H}. In this regime, it reduces to the following relation, with the energy and number density  $u$ and $n$ respectively:
\begin{equation}
P = -u + \mu n\,.
\label{eq:2}
\end{equation}

The single-particle energy $\epsilon$ of a relativistic fermion with mass \(m\) and momentum $p$ is
\begin{equation}
\epsilon(p) = \sqrt{p^2 c^2 + m^2 c^4}= m c^2 \sqrt{1 + a^2}, \qquad a = \frac{p}{m c}\,.
\end{equation}
The non-relativistic (\(a \ll 1\)) and ultra-relativistic (\(a \gg 1\)) limits
recover the familiar expressions \(\epsilon \simeq m c^2 + p^2/(2m)\) and
\(\epsilon \simeq p c\), respectively.

For a system of free fermions occupying the same physical volume, each quantum state occupies a volume \(h^3\) in
momentum space. The number density is therefore given by
\begin{equation}
n = \frac{g}{h^3} \int f_{\mathrm{FD}}(p)\, d^3 p\,.
\end{equation}
Here, $g$ is the degeneracy factor and $f_{\mathrm{FD}}$ the Fermi--Dirac distribution function.

In the zero-temperature limit relevant for compact stars
\begin{equation}
f_{\mathrm{FD}}(\epsilon) =
\begin{cases}
1, & 0<\epsilon \le \mu,\\[4pt]
0, & \epsilon > \mu,
\end{cases}
\end{equation}
and the chemical potential coincides with the Fermi energy $\epsilon_F$ the highest occupied energy state,
\begin{equation}
\mu = \epsilon_F = \sqrt{p_F^2 c^2 + m^2 c^4}\,.
\end{equation}
where $p_F$ is the Fermi momentum.

Assuming spherical symmetry in momentum space
\begin{equation}
n = \frac{g}{(2\pi \hbar)^3} \int_0^{p_F} 4\pi p^2\, dp
  = \frac{g\, p_F^3}{6\pi^2 \hbar^3} ,
  \qquad
  u = \frac{g}{(2\pi \hbar)^3} \int_0^{p_F} 4\pi\, \epsilon(p)\, p^2\, dp\,.
\end{equation}

Introducing the dimensionless Fermi momentum \(x = p_F/(m c)\) and defining
\(E_0 = m^4 c^5/(\pi^2 \hbar^3)\) for spin $1/2$ fermions (\(g=2\)), one finds
\begin{equation}
u = \frac{E_0}{8}\, g(x), \qquad
g(x) = (2x^3 + x)\sqrt{1 + x^2} - \operatorname{arcsinh}(x)
\end{equation}

 Using the thermodynamic identity
Eq.~\eqref{eq:2}, the pressure can be written in compact form as
\begin{equation}
P = \frac{E_0}{24}\, f(x), \qquad
f(x) = (2x^3 - 3x)\sqrt{1 + x^2} + 3\,\operatorname{arcsinh}(x)\,.
\end{equation}

In the non-relativistic limit (\(x \ll 1\)), the leading-order expressions are the following:
\begin{equation}
u_{\mathrm{NR}} = n m c^2
+ \frac{3}{5}\frac{\hbar^2}{2m}(3\pi^2)^{2/3} n^{5/3}, \qquad
P_{\mathrm{NR}} =
\frac{\hbar^2}{5m}(3\pi^2)^{2/3} n^{5/3}\,.
\end{equation}
Whereas, in the ultra-relativistic limit (\(x \gg 1\)) one obtains
\begin{equation}
u_{\mathrm{UR}} =
\frac{3}{4}\hbar c (3\pi^2)^{1/3} n^{4/3}, \qquad
P_{\mathrm{UR}} =
\frac{\hbar c}{4}(3\pi^2)^{1/3} n^{4/3}\,.
\end{equation}

\subsection{Liquid Drop Model}

The liquid drop model, also known as the semi-empirical mass formula, provides a
macroscopic description of nuclear binding energies and remains a useful tool
for modeling nuclear matter in astrophysical environments. While the model
neglects shell effects and therefore cannot reproduce magic numbers or detailed
nuclear structure, it captures the dominant contributions to nuclear binding and
is particularly well suited for describing nuclei with large mass numbers, as
encountered in the crust of neutron stars \cite{2025arXiv250714012F}.

In this semi-classical picture, the nucleus is treated as a droplet of
incompressible nuclear fluid at approximately constant density $\rho_0 \simeq 2.7 \times 10^{14}\,\mathrm{g\,cm^{-3}}$, while
the nuclear radius scales as $R \propto A^{1/3}$ where A is the mass number.
Short-range strong interactions give rise to a bulk (volume) binding energy,
while finite-size effects generate a surface correction.
Long-range Coulomb repulsion between protons contributes an additional positive
energy term.
Kinetic effects associated with the Pauli principle are incorporated
phenomenologically through the volume and asymmetry terms.

The total nuclear energy may be written schematically as \cite{2006PhRvC..73f7302R,1939PhRv...55..963W}:
\begin{equation}
E = Z m_p c^2 + N m_n c^2
+ E_{\mathrm{vol}}
+ E_{\mathrm{surf}}
+ E_{\mathrm{coul}}
+ E_{\mathrm{sym}}
+ E_{\mathrm{pair}}\,
\end{equation}
where $Z$ and $N$ are the proton and neutron numbers, respectively, and of course $A=Z+N$.
$E_{\mathrm{pair}}$ denotes the pairing correction that accounts for odd--even effects in nuclear binding and is typically
small for large mass numbers; it is therefore neglected in the present work.

Introducing the proton fraction $Y_p = Z/A$, the nuclear energy per baryon
can be written in the form
\begin{equation}
\begin{aligned}
\frac{E}{A} =\;&
m_n c^2
+ Y_p (m_p - m_n)c^2
- E_u
+ E_\sigma A^{-1/3}
+ E_c Y_p^2 A^{2/3}
+ E_s (1 - 2Y_p)^2 \,
\end{aligned}
\end{equation}

Here, $E_u$ represents the bulk binding energy per nucleon arising primarily
from the attractive strong interaction.
The surface coefficient $E_\sigma$ accounts for the reduction in binding energy
experienced by nucleons at the nuclear surface.
The Coulomb coefficient $E_c$ quantifies the electrostatic repulsion between
protons and increases with nuclear size.
Finally, the symmetry energy coefficient $E_s$ encodes the energetic cost of
neutron--proton asymmetry, reflecting both kinetic (Fermi energy) and interaction
effects associated with isospin imbalance. 

 Using representative empirical values (in MeV) \cite{2008NuPhA.807..105R}
\begin{equation}
E_u = 15.6, \qquad
E_\sigma = 17.1, \qquad
E_c = 0.71, \qquad
E_s = 23.4\,
\end{equation}

Equilibrium nuclear compositions are obtained by minimizing the total energy
with respect to $A$ and $Y_p$, which reproduce the nuclear curve for stable nuclei.

\subsection{Modeling matter at high densities}

The structure of dense stellar matter is determined by the interplay between
fermionic degeneracy, nuclear binding, and weak-interaction equilibrium.
As density increases, matter undergoes a sequence of well-defined transitions,
each characterized by different dominant degrees of freedom and equilibrium
conditions. In this section we construct the equation of state by treating each
density regime separately, while ensuring thermodynamic consistency across the
transitions.

\subsection*{Ionized matter and electron degeneracy}

Near the surface of compact stars, matter is fully ionized due to high
temperatures. At low densities, electrons behave as a classical ideal gas, while
at higher densities they become degenerate. The electron number density is
\begin{equation}
n_e =Z\frac{\rho}{m_{nuclei}}= \frac{Y_p \rho}{m_u}\,
\end{equation}
where we used $m_{nuclei}=A m_u$ with $m_u$ the average nucleon mass.
 
\subsection*{Neutronization and $\beta$-equilibrium}

At densities $\rho \simeq 10^{8}\,\mathrm{g\,cm^{-3}}$, the electron Fermi
energy (in UR limit) approaches the neutron-proton mass difference, allowing electron capture via the URCA process \cite{1991PhRvL..66.2701L}
\begin{equation}
p + e \rightarrow n + \nu_e, \qquad
n \rightarrow p + e + \bar{\nu}_e\,
\end{equation}
Neutrinos escape the star, so their chemical potential is effectively zero.

Equilibrium is set by
\begin{equation}
\mu_p + \mu_e = \mu_n\,
\label{eq:chem}
\end{equation}
where $\mu_p$, $\mu_e$, and $\mu_n$ are the chemical potentials of protons, electrons, and neutrons, respectively.

The total energy density in this regime is
\begin{equation}
u_{\text{tot}} =
\frac{\rho}{A m_u} E_{\text{nucl}} + \frac{3}{4} n_e \epsilon_{F e}
= \frac{\rho}{m_u}\left(u_{\text{nucl}} + \frac{3}{4} Y_p \epsilon_{F e}\right)\,
\end{equation}
Here, $u_{\text{nucl}}$ is the nuclear energy per baryon, and $\epsilon_{F e}$ the electron Fermi energy.

Equilibrium nuclear configurations are obtained by minimizing $u_{\text{tot}}$:
\begin{equation}
\left(\frac{\partial u_{\text{tot}}}{\partial A}\right)_{Y_p} = 0, \qquad
\left(\frac{\partial u_{\text{tot}}}{\partial Y_p}\right)_{A} = 0\,
\end{equation}
This is equivalent to Eq.~\eqref{eq:chem} and charge neutrality and yields
\begin{equation}
A = \frac{E_\sigma}{2 E_c Y_p^2}\,
\end{equation}
\begin{equation}
(m_n - m_p)c^2 + 4E_s(1 - 2Y_p)
- (2E_\sigma^2 E_c)^{1/3} Y_p^{-1/3}
= \epsilon_{F e}
= \hbar c (3\pi^2)^{1/3}
\left(\frac{Y_p \rho}{m_u}\right)^{1/3}\,
\end{equation}
From the above system, parameters $A$ and $Y_p$ can be solved numerically and for higher densities $A$ increases while $Y_p$ decreases. 
It is concluded that larger nuclei are formed while neutrons are favoured against protons.
\subsection*{Neutron drip}

At $\rho \simeq 4 \times 10^{11}\,\mathrm{g\,cm^{-3}}$ the neutron chemical potential reaches $m_n c^2$ and neutrons become unbound \cite{1971NuPhA.175..225B,2015PhRvC..91e5803C} so we consider the following chemical equilibrium equations 
\begin{equation}
X_Z^A + e \leftrightarrow X^{A}_{Z-1}, \qquad
X_Z^A \leftrightarrow X^{A-1}_Z + n\,
\end{equation}
The first reaction corresponds to electron capture (URCA), while the second describes the emission of free neutrons from nuclei.
We denote $Y_n$ as the fraction of free neutrons over nucleons.
The total energy density is
\begin{equation}
u_{\text{tot}} =
\frac{\rho}{m_u}
\Big(
(1-Y_n)u_{\text{nucl}}
+ \frac{3}{4}(1-Y_n)Y_p \epsilon_{F e}
+ \frac{3}{5} Y_n \bar{\epsilon}_{F n}
+ Y_n m_n c^2
\Big)\,
\end{equation}
with $ \epsilon_{F n}$ the Fermi energy for neutrons and $\bar{\epsilon}_{F n} = \epsilon_{F n} - m_n c^2$ (electrons in UR and free
neutrons in NR limits)

Equilibrium conditions require
\begin{equation}
\left(\frac{\partial u_{\text{tot}}}{\partial A}\right)_{Y_p,Y_n} = 0, \quad
\left(\frac{\partial u_{\text{tot}}}{\partial Y_p}\right)_{A,Y_n} = 0, \quad
\left(\frac{\partial u_{\text{tot}}}{\partial Y_n}\right)_{A,Y_p} = 0\,
\end{equation}
which lead to
\begin{equation}
A = \frac{E_\sigma}{2 E_c Y_p^2}\,
\end{equation}
\begin{equation}
(m_n - m_p)c^2 + 4E_s(1 - 2Y_p)
- (2E_\sigma^2 E_c)^{1/3} Y_p^{-1/3}
= \hbar c (3\pi^2)^{1/3}
\left(\frac{(1-Y_n)Y_p \rho}{m_u}\right)^{1/3}\,
\end{equation}
\begin{equation}
-E_u + E_s(1 - 4Y_p^2)
+ \frac{Y_p^{2/3}}{2}(2E_\sigma^2 E_c)^{1/3}
= \frac{\hbar^2}{2m_n}
(3\pi^2)^{2/3}
\left(\frac{Y_n \rho}{m_u}\right)^{2/3}\,
\end{equation}

\subsection*{Matter at supranuclear densities}

At $\rho \gtrsim 5 \times 10^{13}\,\mathrm{g\,cm^{-3}}$, nuclear clusters
begin to overlap and form complex non-spherical configurations (“nuclear pasta”), rendering the traditional surface and Coulomb energy contributions negligible \cite{2014PhRvC..90f5802P}.

The equilibrium conditions for $Y_p$ and $Y_n$ reduce to
\begin{equation}
(m_n - m_p)c^2 + 4E_s(1 - 2Y_p)
+ \frac{16}{27} E_s (1 - 2Y_p)^3
= \epsilon_{F e}\,
\end{equation}
\begin{equation}
-E_u + E_s(1 - 4Y_p^2)
= m_n c^2 (\sqrt{1 + y^2} - 1)\,
\end{equation}
where
\begin{equation}
y = \frac{p_{F n}}{m_n c}
= \frac{\hbar}{m_n c}
(3\pi^2)^{1/3}
\left(\frac{Y_n \rho}{m_u}\right)^{1/3}\,
\end{equation}

At densities approaching nuclear saturation, $\rho \sim \rho_{0}$, we assume that matter enters a saturation regime characteristic of bound nuclear matter. In this limit, we impose the effective condition
\begin{equation}
Y_n=1-\frac{ \rho_{0}}{{\rho}}\,
\end{equation}
which should be understood as an effective measure of the fraction of matter that actively contributes to pressure support. Near nuclear saturation density, a non-negligible portion of the baryonic content remains locked in correlated or clustered nuclear structures and therefore does not contribute to pressure support. This phenomenological prescription provides a minimal and thermodynamically consistent interpolation between the outer crust and higher-density regimes.
Thus, the specific form adopted for the free neutron fraction $Y_n(\rho)$ is intended as a minimal effective parametrization capturing the onset and growth of neutron drip, rather than a microscopic nuclear calculation. Its role is to interpolate smoothly between the outer crust and higher-density regimes while preserving thermodynamic consistency.
Moderate variations in the functional form and parameters of $Y_n$ will lead only to minor quantitative changes in the resulting equation of state and do not qualitatively affect the global neutron-star properties discussed below. Namely, we have experimented with more generic profiles of the form $Y_{n}=1-\left(\rho_0/\rho\right)^\delta$, within the range $\delta \in [0.8,1.2]$. We found that the variations in the maximum mass for the extreme values of $\delta$ are within $5\%$ compared to the values reported here, corresponding to $\delta=1$.

The pressure arising solely from fermionic degeneracy is generally insufficient to ensure mechanical stability. It is therefore customary to supplement the equation of state with an additional polytropic pressure component.

Polytropic equations of state play a central role in astrophysics and have been extensively employed in the modeling of self-gravitating systems, including main-sequence stars, white dwarfs, neutron stars, and core-collapse simulations \cite{1965ApJ...142.1541T,1965PhDT.........4T}. Their widespread use stems from their ability to capture the essential macroscopic behaviour of matter under compression while remaining analytically and numerically tractable.
The additional polytropic pressure is taken to be
\begin{equation}
P_{polytrop} = K \left( \frac{Y_n\,\rho}{\rho_{0}} \right)^{\gamma}\,
\label{eq=pol}
\end{equation}
where $K$ is the polytropic constant and $\gamma$ denotes the adiabatic index.

Mesonic polytropes are the best candidates for a polytropic description of the equation of state
at supranuclear densities, where short-range strong interactions between baryons become dominant
and the free relativistic Fermi-gas approximation must be extended.
We adopt a Quantum Hadrodynamics (QHD) mean-field approximation (MFA) framework, in which nucleons interact via
effective meson exchange \cite{1974AnPhy..83..491W,1997IJMPE...6..515S}.

The interaction Lagrangian is given by
\begin{equation}
\mathcal{L}_{\mathrm{int}}
=
\bar{\psi}
\left[
- g_\sigma \sigma
- g_\omega \gamma^\mu \omega_\mu
- g_\rho \gamma^\mu \vec{\tau}\cdot\vec{\rho}_\mu
\right]\psi
+ \mathcal{L}_{\sigma}
+ \mathcal{L}_{\omega}
+ \mathcal{L}_{\rho}\,
\end{equation}
where $g_i$,$m_i$ and $\mathcal{L}_{i}$ are the strength, the mass and the free Lagrangian of each meson. 
In uniform infinite matter, the meson fields are replaced by their mean-field expectation values (MFA).
Neglecting nonlinear scalar self-interactions at this stage, the mesonic contribution to the pressure
takes the form 
\begin{equation}
P_{\mathrm{mes}}
=
- \frac{1}{2} m_\sigma^2 \sigma^2
+ \frac{1}{2} m_\omega^2 \omega_0^2
+ \frac{1}{2} m_\rho^2 \rho_{0,3}^2 
\end{equation}
The scalar field $\sigma$ reduces the neutron effective mass
\begin{equation}
m_n^\ast = m_n - g_\sigma \sigma\,
\end{equation}
In relativistic mean-field approaches, nonlinear self-interaction terms of the scalar $\sigma$ field are introduced phenomenologically to reproduce nuclear saturation properties. Such terms, however, are not required by underlying symmetries and mainly serve as effective parametrizations of many-body dynamics. In the present work, we adopt a minimal mean-field description in which the dominant nonlinear scalar effects are absorbed into a density-dependent effective nucleon mass. This choice preserves a particularly transparent structure for the mesonic pressure contributions, which remain quadratic in the relevant densities and naturally resemble a polytropic form. 
Thus, the effective mass is parametrized phenomenologically by a saturating form inspired by relativistic-mean-field  calculations 
\cite{1997BrJPh..27..342D}.
\begin{equation}
\frac{m_n^\ast}{m_n}
=
m_\infty
+
\frac{1 - m_\infty}{1 + C_s \left( Y_n \rho / \rho_0 \right)^{2/3}} \,
\end{equation}
Here, $m_\infty$ denotes the finite asymptotic high density limit of the effective nucleon mass, while $C_s$ controls the rate of scalar-density suppression. 
The exponent $2/3$ is motivated by Fermi-momentum scaling, $k_F \propto n^{1/3}$, such that $(\rho/\rho_0)^{2/3}$ effectively tracks a $k_F^2$-type density dependence, commonly appearing in relativistic mean-field and microscopic descriptions of dense Fermi systems. 
This form ensures the correct low-density limit, $m_n^\ast/m_n \to 1$ as $\rho \to 0$, while allowing for a smooth decrease and saturation of the effective mass at supranuclear densities. 
Values $m_\infty \simeq 0.45$--$0.55$, $C_s=2$ are consistent with standard RMF models for effective mass in nuclear densities and above \cite{2004BrJPh..34..833S,2016NuPhA.950...64M}.

The mean-field equations (Euler-Lagrange) yield
\begin{equation}
\sigma_0 = \frac{g_\sigma}{m_\sigma^2} n_s,
\qquad
\omega_0 = \frac{g_\omega}{m_\omega^2} n,
\qquad
\rho_{0,3} = \frac{g_\rho}{m_\rho^2} n,
\end{equation}
where $n$ is the number density of free neutrons and $n_s$ is the scalar number density; 
the scalar density is evaluated analytically using the relativistic Fermi integral,
\begin{equation}
n_s
=
\frac{m_n^{\ast 3} c^3}{\pi^2 \hbar^3}
\left[
y \sqrt{1+y^2}
- \ln \left( y + \sqrt{1+y^2} \right)
\right],
\qquad
y = \frac{p_F}{m_n^\ast c}\,
\end{equation}

Introducing the effective couplings $G_i$, the total mesonic pressure in the $\sigma \omega\rho$ model can be written as
\begin{equation}
P_{\mathrm{mes}}
=
- \frac{1}{2} G_\sigma n_s^2
+ \frac{1}{2} G_\omega n^2
+ \frac{1}{2} G_\rho n^2 ,
\qquad
G_i \equiv \frac{g_i^2}{m_i^2}\,
\end{equation}

The total neutron pressure in the inner core is finally written as
\begin{equation}
P(\rho, Y_n)
=
P^{\mathrm{^\ast}}_{n}(\rho,Y_n)
+
P_{\mathrm{mes}} (\rho,Y_n)\,
\end{equation}
where $P^{\mathrm{^\ast}}_{n}$ denotes the full relativistic Fermi-gas pressure
evaluated with the density-dependent effective mass.

The effective couplings $G_i$ introduced above are not fundamental constants, but rather
model-dependent parameters that encode both the meson masses and their corresponding
interaction strengths.

For the vector mesons $\omega$ and $\rho$, the physical masses are well established
experimentally. However, their effective coupling strengths $g_\omega$ and $g_\rho$
are determined phenomenologically and depend on the specific RMF parameterization
employed (e.g., NL3, TM1, PK1, DD-ME2), leading to corresponding variations in $G_\omega$
and $G_\rho$ across different models.

The situation is more subtle for the scalar channel. The $\sigma$ field does not represent
a well-defined physical particle, but rather an effective degree of freedom that
parametrizes correlated two-pion exchange and intermediate-range attraction between
nucleons. As a consequence, the scalar ``mass'' $m_\sigma$ is not uniquely defined,
and both $g_\sigma$ and $m_\sigma$ should be regarded as effective parameters.

It follows that the couplings $G_i$ cannot be directly compared across different RMF
models without accounting for differences in the treatment of scalar dynamics,
effective masses, and density dependence. In the present work, the $G_i$ are therefore
interpreted as effective, renormalized couplings, calibrated within a given mean-field
framework rather than as universal quantities.

\subsection*{Summary of pressure contributions}

Across all density regimes, the number densities of free fermions are
\begin{equation}
n_e = (1 - Y_n) Y_p \frac{\rho}{m_u}, \qquad
n_n = Y_n \frac{\rho}{m_u}\,
\end{equation}
and the dimensionless Fermi momenta are
\begin{equation}
x = \frac{p_{F e}}{m_e c}
= \frac{\hbar}{m_e c}(3\pi^2)^{1/3} n_e^{1/3}, \qquad
y = \frac{p_{F n}}{m_n^\ast c}
= \frac{\hbar}{m_n^\ast c}(3\pi^2)^{1/3} n_n^{1/3}\,
\end{equation}

The total pressure is obtained from
\begin{equation}
P = \sum_i \mu_i n_i - u_{\text{tot}}=P_e (\rho,Y_p,Y_n)+P^{\mathrm{^\ast}}_{n}(\rho,Y_n)+P_{mes}\,
\label{eq:pressure_decomposition}
\end{equation}
The contributions depend on the dominant degrees of freedom in each density regime. At low densities, pressure is dominated by degenerate electrons; beyond neutron drip, free neutrons contribute a non-relativistic Fermi pressure. At supranuclear densities, $P$ is mainly determined by the polytropic term or by the meson pressure in the RMF approximation.

At each transition between density regimes, the equation of state is constructed to ensure continuity of the pressure and energy density. The matching procedure enforces smooth thermodynamic behaviour across the boundaries, so that no unphysical discontinuities are introduced in the stellar structure calculations.

\subsection{Relativistic Hydrostatic Equilibrium}

The global structure of a neutron star is determined by the balance between
pressure gradients and gravity. In the relativistic regime relevant for compact
objects, this balance is governed by Einstein’s field equations,
\begin{equation}
{\cal R}_{\mu\nu} - \frac{1}{2} {\cal R} g_{\mu\nu}
= -\frac{8\pi G}{c^4} T_{\mu\nu}\,.
\end{equation}
Here, ${\cal R}_{\mu\nu}$ is the Ricci tensor, ${\cal R}$ the Ricci scalar, $g_{\mu\nu}$ the metric tensor, $G$ the gravitational constant and $T_{\mu\nu}$ the energy--momentum tensor.

For a static, spherically symmetric configuration, matter is modelled as a perfect fluid:
\begin{equation}
T_{\mu\nu}
= \left( \rho + \frac{P}{c^2} \right) u_\mu u_\nu - P g_{\mu\nu}\,.
\end{equation}
Symbols $\rho$ and $P$ denote the mass--energy density and pressure. The four-velocity $u^\mu$ satisfies the normalization condition $u^\mu u_\mu = c^2$ and, 
in the local rest frame of the fluid it has only a temporal component corresponding to a static configuration.

The enclosed gravitational mass within radius $r$ is defined as
\begin{equation}
m(r) = \int_0^r 4\pi r'^2 \rho(r')\, dr'\,.
\end{equation}

The condition of hydrostatic equilibrium takes the form
\begin{equation}
\frac{dP}{dr}
= -\frac{G}{r^2}
\left( \rho + \frac{P}{c^2} \right)
\left( m(r) + \frac{4\pi P r^3}{c^2} \right)
\left( 1 - \frac{2G m(r)}{c^2 r} \right)^{-1}\,.
\end{equation}
This is the Tolman--Oppenheimer--Volkoff (TOV) equation, describing the balance of forces in a relativistic star \cite{1939PhRv...55..364T,1939PhRv...55..374O}.

The TOV equation reduces to the familiar Newtonian hydrostatic equilibrium equation in the weak-field limit ($P \ll \rho c^2$, $2Gm/(c^2 r) \ll 1$), but relativistic corrections are essential for neutron stars, strongly affecting the maximum mass and the mass--radius relation.

To construct a stellar model, the TOV equations are supplemented by an equation of state $P(\rho)$, a central pressure $P_c$, and the surface boundary condition $P(R)=0$, which defines the stellar radius $R$. Numerical integration from the center outward determines pressure, density, and enclosed mass, yielding the total gravitational mass
\begin{equation}
M = m(R)\,.
\end{equation}
Physical constraints impose additional limits. The adiabatic sound speed $c_s$ must be less than the speed of light to satisfy causality. For each equation of state considered, the adiabatic sound speed 
\begin{eqnarray}
c_s = \sqrt{\frac{\partial P}{\partial \rho}}
\end{eqnarray}
was evaluated numerically. Only EOSs satisfying the causality condition $c_s<c$, over the entire density range were retained for the TOV integrations presented below. Moreover, the total gravitational mass $M$ must increase with the increase of central pressure:
\begin{equation}
\frac{dM}{dP_c} > 0\,.
\end{equation}
This is a necessary but not sufficient condition for the radial stability of the star \cite{1964ApJ...140..417C}. From Bardeen-Thorne-Meltzer (BTM) criteria \cite{1965gtgc.book.....H}, it is concluded that configurations beyond the maximum of the $M-R$ curve are unstable in our case.

Once a solution of the TOV equations is obtained, macroscopic quantities such as the stellar radius, gravitational binding energy, compactness, surface redshift, and moment of inertia can be extracted. These provide a direct link between microphysics encoded in the equation of state and observable neutron-star properties.

The compactness is defined as follows
\begin{equation}
\beta = \frac{G M}{c^2 R}\,.
\end{equation}
Two important measures of relativistic effects are the surface gravitational redshift $z$ and the apparent radius $R_\infty$
\begin{equation}
1 + z = (1 - 2\beta)^{-1/2}\,,
\qquad
R_\infty = (1 + z) R\,,
\end{equation}
where $R_\infty$ is essentially the radiation radius measured by a distant observer which exceeds the physical circumferential radius due to spacetime curvature.  This distinction is essential in interpreting thermal X-ray spectra, where 
observables involve \(R_\infty\) rather than \(R\).

\section{Results for Neutron Stars}

\label{sec:neutron_stars}

Using the equation of state constructed in the previous sections, the Tolman--Oppenheimer--Volkoff (TOV) equations were solved numerically for a wide range of central pressures.
The integration was performed using a Runge--Kutta--Fehlberg (RKF 4/5) scheme, while the proton and neutron fractions,
$Y_p$ and $Y_n$, were determined self-consistently at each density point via a Newton--Raphson iteration.  Convergence was verified by varying the integration tolerance and step size, with global stellar properties changing by less than $1\%$. 
We note that the TOV solutions are highly sensitive to the adopted microscopic parameters (e.g.\ effective couplings and saturation inputs),
and large deviations from physically motivated values may lead to nonphysical stellar configurations.

The stellar surface was defined by the standard condition
$P(R)=0$, while the surface temperature was fixed to
$T_{\mathrm{surf}} = 10^7\,\mathrm{K}$.
Central pressures were scanned in the interval
\begin{equation}
P_c \in [10^{34},\,10^{36}]\,\mathrm{dyn\,cm^{-2}}\,,
\end{equation}
and only stable solutions satisfying the condition $dM/dP_c > 0$ were retained.

\subsection{Observational data}

We will confront our theoretical results with a set of observational constraints derived from well-studied neutron stars. 
The most direct mass--radius measurements are provided by the \textit{NICER} mission through X-ray pulse profile modeling. 
In particular, PSR~J0030+0451 \cite{2025arXiv251208790M} and PSR~J0740+6620 \cite{2021ApJ...918L..27R} yield simultaneous constraints on mass and radius, while the nearby millisecond pulsar PSR~J0437--4715 \cite{2021ApJ...918L..29R} provides an additional high-quality radius measurement.
These objects currently constitute the most reliable direct probes of the neutron-star mass--radius relation.

Independent constraints on the maximum supported mass are obtained from radio timing observations of massive pulsars, notably PSR~J0348+0432 \cite{2013Sci...340..448A}, PSR~J1614--2230 \cite{2010Natur.467.1081D}, and PSR~J0952--0607 \cite{2022ApJ...934L..17R}, note however that the latter is the fastest rotating known neutron star. However, our solutions are static and non-rotating. 
Although no radius information is available for these systems, their accurately measured masses above $2\,M_\odot$ impose strong lower bounds on the stiffness of the equation of state.

Additional information is available for the gravitational redshift and the apparent radius. 
A surface gravitational redshift $z \simeq 0.35$ has been reported for the low-mass X-ray binary EXO~0748--676 \cite{2002xrb..confE...5C}. While this remains the only direct spectroscopic
redshift measurement reported for a neutron-star surface, its interpretation
has been subject to debate and has not been independently confirmed.

Apparent radii $R_\infty$ have been inferred for several neutron stars, including NICER pulsars and quiescent neutron stars in globular clusters, such as X7 in 47~Tuc \cite{2013AAS...22141201G,2016ApJ...831..184B} .
These observational constraints are summarized in Table~\ref{tab:observations} and are used to assess the consistency of equations of state with current astrophysical data \cite{2016ARA&A..54..401O}.
\begin{table}[h!]
\centering
\caption{Summary of observational constraints used in this work.}
\label{tab:observations}
\begin{tabular}{lcccc}
\hline
Object & $M\,[M_\odot]$ & $R$ [km] & $z$ & $R_\infty$ [km] \\
\hline
PSR J0030+0451 & $1.34\text{--}1.44$ & $12.7\text{--}13.0$ & $\sim0.20$ & $\sim16$ \\
PSR J0740+6620 & $2.08\pm0.07$ & $12.9\text{--}13.7$ & $\sim0.35$ & $\sim18$ \\
PSR J0437--4715 & $1.418\pm0.037$ & $11.3\text{--}11.5$ & -- & $14\text{--}15$ \\
PSR J0348+0432 & $2.01\pm0.04$ & -- & -- & -- \\
PSR J1614--2230 & $1.908\pm0.016$ & -- & -- & -- \\
PSR J0952--0607 & $2.35\pm0.17$ & -- & -- & -- \\
EXO 0748--676 & $\sim1.4$ & -- & $0.35$ & -- \\
X7 (47 Tuc) & $\sim1.4$ & -- & -- & $14.3\pm2.5$ \\
\hline
\end{tabular}
\end{table}

\subsection{Single-polytrope analysis}

The short-range repulsive interaction introduced at supranuclear densities should be interpreted as the effective manifestation of underlying QCD dynamics, rather than as an ad hoc addition. In phenomenological descriptions of dense matter, such repulsion is universally encoded through effective interactions, for example via vector meson exchange in relativistic mean-field models \cite{Glendenning:2000, Haensel:2007}. In the present framework, this repulsive component plays the minimal role required to ensure stability and causality of matter at high densities, without introducing additional microscopic assumptions \cite{Lattimer:2007}.

\begin{figure}
  \centering
  \includegraphics[width=0.8\linewidth]{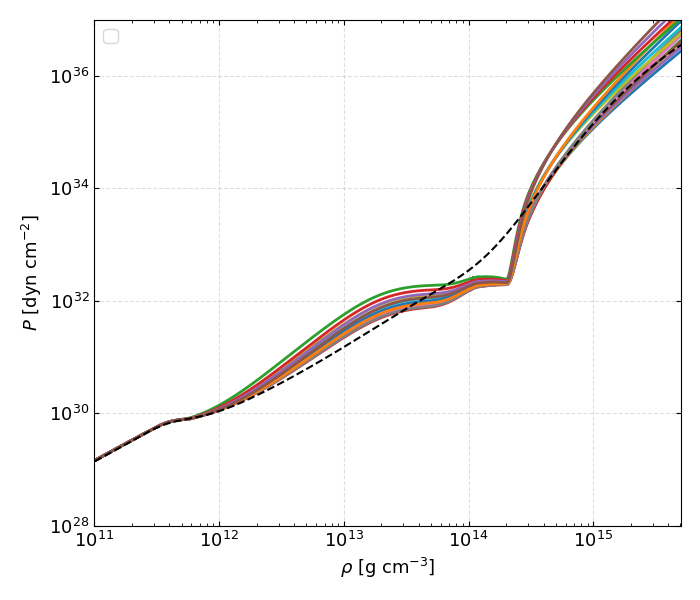}
  \caption{Pressure--density relation at high densities for a one-term polytropic equation of state
  with different values of $K$ and $\gamma$. Each curve is compared against the SLy equation of state,
  which serves as a benchmark for realistic nuclear matter behaviour.}
  \label{fig:P-n (K)}
\end{figure}

\begin{figure}
  \centering
  \includegraphics[width=0.8\linewidth]{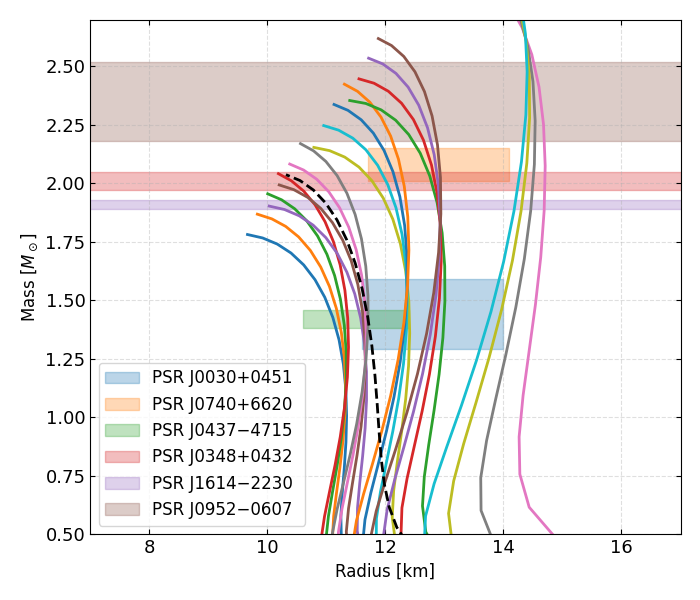}
  \caption{Mass--radius relations obtained for multiple $(K,\gamma)$ combinations.
  Observational constraints are indicated for comparison, together with the SLy4 prediction.}
  \label{fig:M-R (K)}
\end{figure}

As a first step, we examined a single-polytrope parametrization of the high density equation of state,
using Eq.~\eqref{eq:pressure_decomposition} and scanning a wide range of polytropic constants $K$ and polytropic indices $\gamma$.
This exploratory analysis allows us to identify combinations of parameters capable of reproducing
realistic pressure-density relation Fig.~\ref{fig:P-n (K)} and neutron-star masses and radii Fig.~\ref{fig:M-R (K)} - we compare with Sly4 EoS \cite{2004A&A...428..191H}. The non-smooth behavior visible in some pressure–density curves reflects the exploratory nature of the single-polytrope scan and the matching to lower-density prescriptions, rather than a physical phase transition. 

The resulting mass--radius sequences exhibit a well-defined maximum mass, typically in the range
$M_{\max} \simeq 1.5$--$3\,M_\odot$, depending on the stiffness of the equation of state.
For each parameter set, the stable branch extends from $M \sim 1\,M_\odot$ up to the maximum mass,
beyond which the sequence terminates abruptly, signalling the onset of relativistic instability as dictated by the TOV equations.

\begin{figure}
  \centering
  \includegraphics[width=0.8\linewidth]{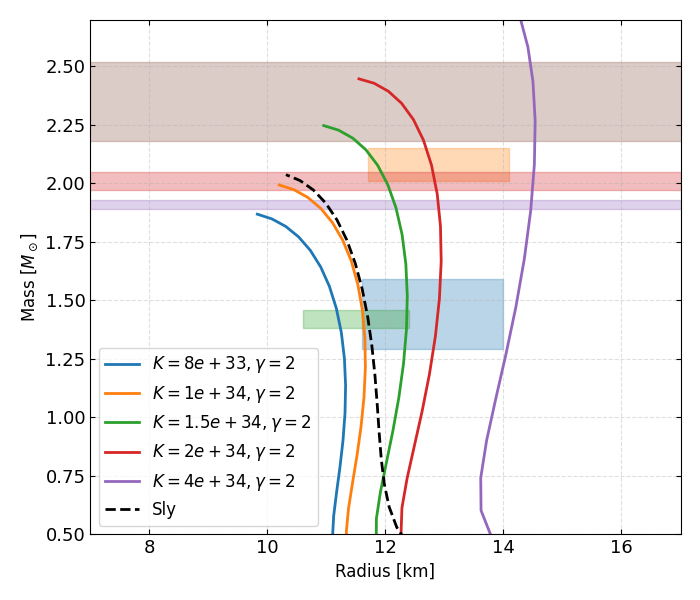}
  \caption{Mass--radius relations for the subset of polytropic equations of state that satisfy
  observational constraints and yield physically viable neutron stars.}
  \label{fig:M-R (K)!}
\end{figure}

It is evident that a large fraction of the explored parameter space does not yield neutron-star
configurations compatible with current observational constraints.
Realistic solutions are found only for a narrow subset of parameters, characterized by a
polytropic index $\gamma \simeq 2$ and stiffness values $K$ close to $10^{34}$ in cgs units Fig.~\ref{fig:M-R (K)!}.

\subsection{Mapping polytropes to mesonic couplings}

Having constrained the acceptable range of $K$, we proceeded to relate the effective polytropic
description to the mesonic mean-field framework.
Assuming comparable scalar and baryon densities at high density, $n_s \simeq n = Y_n \rho / m_n$,
the mesonic pressure contribution in Eq.~\eqref{eq:pressure_decomposition} can be approximated by
\begin{equation}
\frac{1}{2}
\left(
- G_{\sigma}
+ G_{\omega}
+ G_{\rho}
\right)
n^2
=
K
\left(
\frac{Y_n \rho}{\rho_0}
\right)^2\,. 
\end{equation}
For representative values $K \simeq 1.5 \times 10^{34}$ in the cgs system, the above relation implies
\begin{equation}
- G_{\sigma}
+ G_{\omega}
+ G_{\rho}
\simeq
1.1 \times 10^{-42} \, .
\end{equation}
These scales are consistent with values commonly encountered in relativistic mean--field models,
which typically yield effective couplings $G_i \sim 10^{-42}$
\cite{2012wmnp.confE..58F,2021PhRvC.104e5804K}.
Since the scalar contribution appears with a negative sign, relatively smaller values of
$G_\sigma$ are favoured to ensure mechanical stability.

Rather than adopting a single parameter choice, in the present work we explored
three representative sets of coupling constants, denoted as G1, G2 and G3,
which all satisfy the above constraint while spanning a reasonable range of
repulsive and isovector contributions.
The adopted values, expressed in cgs units, are summarized in Table~\ref{tab:Gsets}.

\begin{table}[h]
\centering
\begin{tabular}{c c c c}
\hline
Set & $G_{\sigma}$  &
      $G_{\omega}$ &
      $G_{\rho}$    \\
\hline
SET G1 & $2.0 $ & $10.0$ & $9.0 $ \\
SET G2 & $4.0 $ & $11.0 $ & $8.0 $ \\
SET G3 & $1.0 $ & $7.0 $  & $7.0 $ \\
\hline
\end{tabular}
\caption{Representative sets of effective RMF couplings explored in this work in $10^{-43}\mathrm{cm^3\,erg^{-1}}$ values.}
\label{tab:Gsets}
\end{table}

\subsection{Mesonic equation of state and stellar properties}

\begin{figure}
  \centering
  \includegraphics[width=0.8\linewidth]{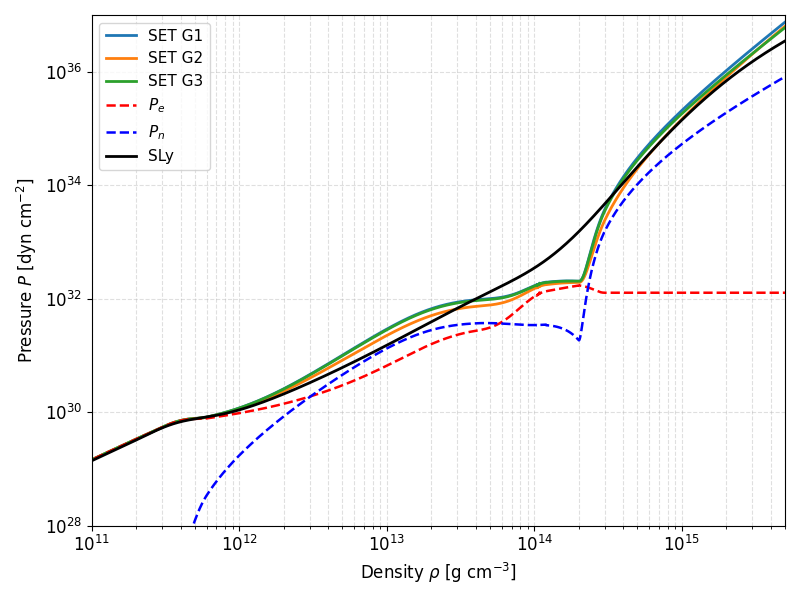}
  \caption{Pressure as a function of density for the adopted set of mesonic couplings $G_i$,
  compared with the SLy4 equation of state. The individual contributions from neutrons and electrons in the
relativistic Fermi gas are shown separately, illustrating the expected behaviour across different
density regimes.
  }
  \label{fig:P-ρ (G)}
\end{figure}

We used this set of effective couplings and the full mesonic pressure contribution, constructed the complete equation of state and solved the TOV equations.
The resulting pressure--density relation is shown in Fig.~\ref{fig:P-ρ (G)}.

\begin{figure}
  \centering
  \includegraphics[width=0.8\linewidth]{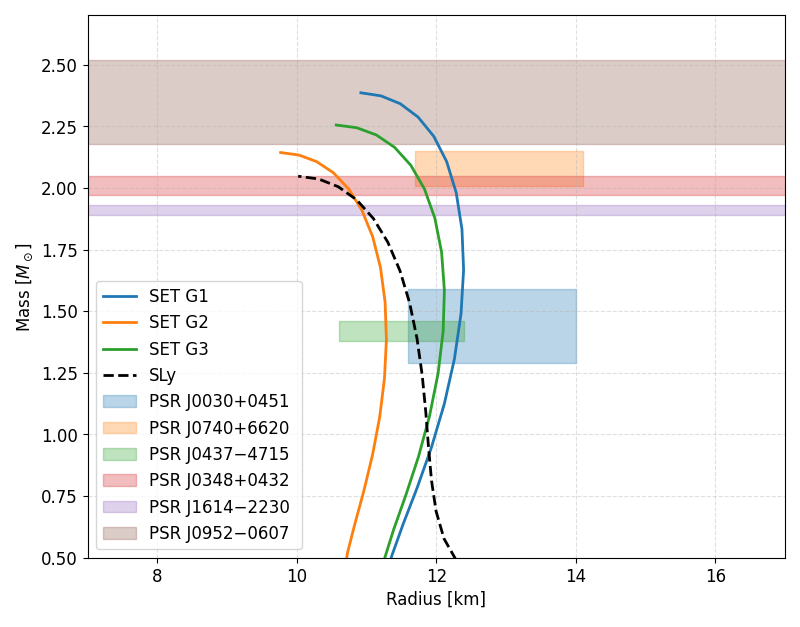}
  \caption{Mass--radius diagram obtained within the $\sigma\omega\rho$ meson-exchange model
  in the mean-field approximation.
  The SLy4 result is included for direct comparison.}
  \label{fig:M-R (G)}
\end{figure}

The corresponding mass--radius relation is displayed in Fig.~\ref{fig:M-R (G)}.

\begin{figure}
  \centering
  \includegraphics[width=0.8\linewidth]{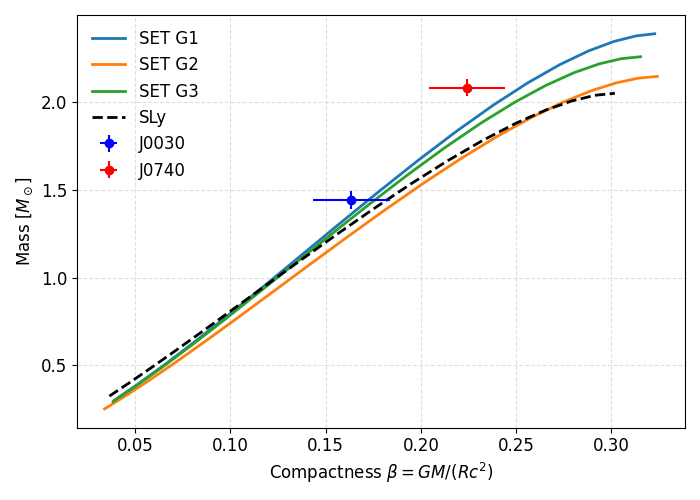}
  \caption{Mass--compactness relation, compared with current observational constraints.}
  \label{fig:beta}
\end{figure}

\begin{figure}
  \centering
  \includegraphics[width=0.8\linewidth]{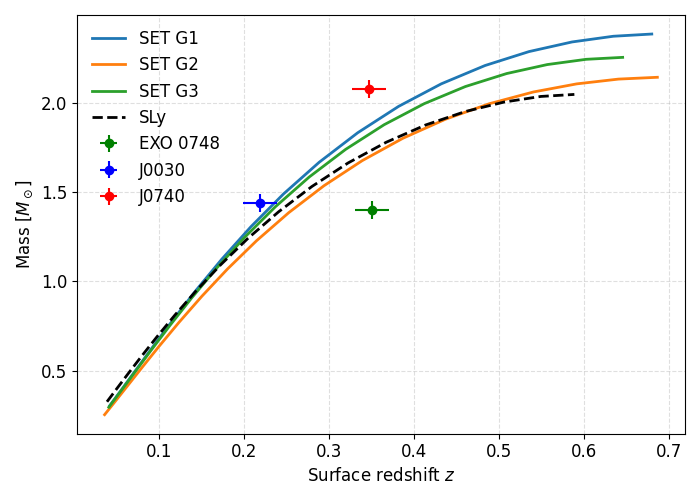}
  \caption{Gravitational redshift as a function of stellar mass.}
  \label{fig:z}
\end{figure}

\begin{figure}
  \centering
  \includegraphics[width=0.8\linewidth]{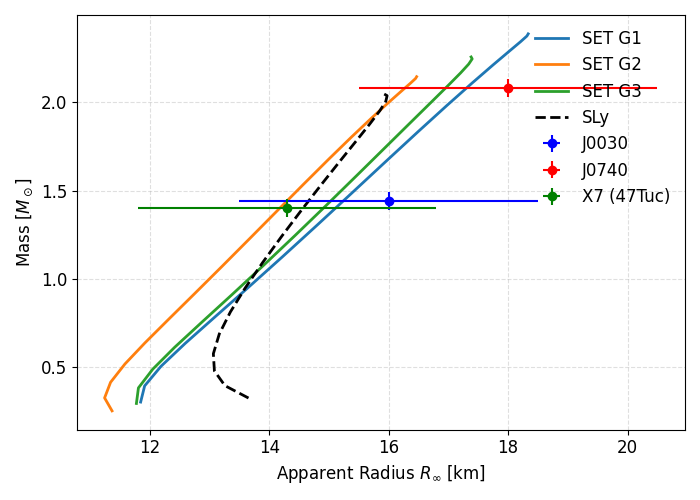}
  \caption{Mass--apparent radius relation for the adopted mesonic equation of state.}
  \label{fig:Rapp}
\end{figure}

Once a consistent solution of the TOV equations has been obtained for a given equation of state,
additional macroscopic neutron-star observables can be examined Figs.~\ref{fig:beta}-\ref{fig:Rapp}.
All the above figures include SLy4 as a reference for visual comparison.

\section{Conclusions}
\label{sec:Conclusions}
In this work, we have developed a simple and physically transparent framework for
modeling neutron-star structure across a wide range of densities. By combining
the statistical mechanics of degenerate fermions with the liquid drop model of
nuclear binding, we constructed an equation of state that smoothly connects the
outer crust, inner crust, and core of a neutron star. The inclusion of a
phenomenological repulsive interaction at high density ensures stability and
causality, and also enabled us to calibrate parameters for more realistic meson pressure terms, which were then used consistently in the
Tolman--Oppenheimer--Volkoff equations.

A qualitative insight emerging from this study is that many macroscopic neutron-star properties---such as masses, radii, compactness, and surface redshifts---are primarily sensitive to the overall stiffness of the equation of state, rather than to detailed microscopic composition. Within a minimal and physically consistent framework, current observational constraints largely probe effective thermodynamic properties of dense matter, while leaving significant degeneracy in the underlying QCD microphysics. This delineates both the power and the limitations of neutron-star observations in constraining the behaviour of matter at supranuclear densities.

Despite the simplicity of the underlying microphysics, the resulting neutron-star
models reproduce key qualitative and quantitative features expected from more
sophisticated approaches. The predicted mass--radius relation, maximum mass,
 surface redshift and apparent radius all fall
within ranges compatible with current astrophysical constraints.

An important advantage of the present approach is its conceptual clarity and
computational simplicity. The equation of state is constructed from a few physically motivated ingredients, making it straightforward to analyse
and numerically stable to implement. This simplicity also facilitates systematic
extensions of the model. Additional particle species, such as muons, can be
included by introducing their corresponding energy density and pressure
contributions, together with an additional chemical potential and an equilibrium
condition. More generally, new degrees of freedom or interaction terms can be
incorporated in a controlled manner by extending the energy density and deriving
the corresponding pressure self-consistently.

Such extensions provide a natural path toward more realistic descriptions of
dense matter while retaining the transparency of the underlying physics. The
framework presented here therefore serves not only as a useful and clear model,
but also as a flexible baseline for exploring the impact of additional physical
effects on neutron-star structure. In this sense, simple models of dense matter
remain a valuable tool for building intuition and for bridging the gap between
microphysical assumptions and observable astrophysical phenomena.

\authorcontributions{ Conceptualization, E.A. and K.N.G.; writing---original draft preparation, E.A.; writing---review and editing, E.A. and K.N.G. The authors have read and agreed to the published version of the manuscript.}

\funding{KNG acknowledges funding from grant FK 81641 "Theoretical and Computational Astrophysics", ELKE. This work was supported by computational time granted by the National Infrastructures for Research and Technology S.A. (GRNET S.A.) in the National HPC facility - ARIS - under project ID pr017008/simnstar2.}

\dataavailability{Numerical data related to this study are available upon request.} 

\acknowledgments{The authors are grateful to Dimitrios Zoakos, Carlos A.J.P. Martins and Eleanna Kolonia for insightful discussions.}

\conflictsofinterest{The authors declares no conflict of interest. The funders was
not involved in the study design, collection, analysis, interpretation
of data, the writing of this article or the decision to submit it for
publication.}

\end{paracol}

\reftitle{References}


\externalbibliography{yes}
\bibliography{Bibtex1.bib}


%


\end{document}